# Direction dependent thermal conductivity of monolayer phosphorene: parameterization of Stillinger-Weber potential and molecular dynamics study

Wen Xu,[1,2] Liyan Zhu,[3] Yongqing Cai,[2] Gang Zhang,[2,*] and Baowen Li[1,4,5]

[1]Department of Physics, Centre for Advanced 2D Material and Centre for Computational Science and Engineering, National University of Singapore, Singapore 117546, Singapore

[2] Institute of High Performance Computing, A*STAR, Singapore 138632, Singapore

[3] Department of physics, Huai'an Normal University, Jiangsu, 223000, China

[4] NUS Graduate School for Integrative Sciences and Engineering, National University of Singapore, Kent Ridge 119620, Singapore

[5] Center for Phononics and Thermal Energy Science, School of Physics Science and Engineering, Tongji University, 200092 Shanghai, China

* E-mail: zhangg@ihpc.a-star.edu.sg

**Abstract**

A Stillinger-Weber interatomic potential is parameterized for phosphorene. It well reproduces the crystal structure, cohesive energy and phonon dispersion predicted by first-principles calculations. The thermal conductivity of phosphorene is further explored by equilibrium molecular dynamics simulations adopting the optimal set of potential parameters. At room temperature, the intrinsic thermal conductivities along zigzag and armchair directions are about 152.7 and 33.0 W/mK, respectively, with a large anisotropy ratio of five. The remarkably directional dependence of thermal conductivity in phosphorene, consistent with previous reports, is mainly due to the strong anisotropy of phonon group velocities, and weak anisotropy of phonon lifetimes as revealed by lattice dynamics calculations. Moreover, the effective phonon mean free paths at zigzag and armchair directions are about 141.4 and 43.4nm, respectively.

# I. INTRODUCTION

With the continuous downscaling of electronic devices, short channel effects have severely affected their performance. Fortunately, the two-dimensional(2D) semiconductors, such as $MoS_2$[1], are immune to these effects, which makes 2D semiconducting nanosheets receive a significant amount of attention. Most recently, another novel 2D semiconductor, monolayer phosphorene, which exhibits fascinating physical properties including a sizable direct band gap, high carrier mobilities, and a large on-off current ratio[2, 3], has been exfoliated. These superior electronic properties render phosphorene a promising candidate in many nanoelectronic and optoelectronic applications.

In addition to the electronic and optical properties, thermal properties of nano materials have also attracted considerable attention due to their unique features different from their counterpart in macroscale[4]. On one hand, thermal management in nanoscale devices with a high power density is a critical issue, where a high thermal conductivity material is needed to dissipate the Joule heat as quickly as possible. On the other hand, phosphorene has been predicted to be a potential thermoelectric material[5, 6], where a low thermal conductivity is desired to achieve a high efficiency. With either motivation, it is necessary to perform a thorough study on the thermal transport properties of phosphorene.

Since phosphorene is a semiconductor, the dominant thermal carriers in it are phonons. So far, theoretical assessment of phonon transport in phosphorene has been done with nonequilibrium Green's function (NEGF) method[7] or by solving the Boltzmann transport equation (BTE)[8, 9], while little is related to molecular dynamics (MD) simulations. MD simulation is another powerful tool to handle many-body problems at the atomic level. It approaches the thermal transport problems without the thermodynamic-limit assumption, and naturally includes full anharmonicity in atomistic interactions. Besides predicting the thermal conductivity, the MD simulation is also very efficient in studying problems such as interfacial thermal resistance[10], thermal rectification[11], mechanical properties[12] and fractural process[13] of nano materials. The applications of MD simulations have covered a wide range of research systems, such as liquids, clusters, and biomolecules[14]. In MD simulations, the force on each atom according to the force field is calculated to numerically solve the Newton's equations of motion. Although a fully quantum-mechanical treatment of the system's Hamiltonian is highly desirable, it can only be applied to the system with a small size due to the large computational load. Therefore, tremendous efforts have been devoted to developing empirical potential fields that can be applied to a large system with the sacrifice of exactness. However, so far there is no existing empirical interatomic potential for monolayer phosphorene yet. Therefore, it is the demand to study thermal transport in phosphorene, the power of MD simulations and the lack of empirical interatomic potential for phosphorene that inspire our work reported in this paper.

In this work, we use quantities obtained from first-principles calculations to parameterize an empirical potential for phosphorene, which is a usual strategy and has been applied to investigate thermal transport in various materials[15-17]. Specifically, the well-established Stillinger-Weber (SW) potential[18] is used as the prototype for fitting, due to its simple form and wide applications[17, 19, 20]. The parameterized potential well reproduces the crystal structure, cohesive energy and phonon dispersion of phosphorene predicted by first-principles calculations. Subsequently, MD simulations are performed with this potential to study the thermal transport in phosphorene. Good agreement between results in this work and previously reported results is reached. This paper is organized in the following way: In Sec. II, the parameterization of SW potential is described and the final parameter set is reported. In Sec. III, the thermal conductivity of phosphorene is calculated with equilibrium MD simulations. In Sec. IV, the phonon properties such as group velocities, lifetimes and mean free paths are evaluated. Finally, in Sec. V, deficiencies of the present SW potential and future developing directions are discussed, followed with a summary of this paper.

## II. PARAMETERIZATION OF POTENTIAL

Phosphorene consists of a sheet of phosphorus atoms puckered along the so-called armchair direction, with four atoms in each unit cell (see Figure **1**). Noticing that the phosphorus atoms occupy two planes parallel with XY plane, we enumerate them with P1 and P2 accordingly. There are two nonequivalent types of P-P bonds, one is parallel, and the other is unparallel with XY plane. Similarly, there are two types of bond angles. We distinguish the two types of bonds and bond angles with subscripts 'in' and 'out' (see Figure **1**(b)). The SW potential is expected to properly describe interactions at these bonds and bond angles. It is worth mentioning that P1 and P2 are just used for specifying the interatomic interactions as will be shown. According to first-principles calculations[8], the lattice constants at X and Y directions are 3.301Å and 4.601Å respectively. The other geometrical quantities are as follows: $l_{in}$ = 2.221 Å, $l_{out}$ = 2.259 Å, $\theta_{in}$ = 96.001°, and $\theta_{out}$ = 103.961°.

The SW potential was initially proposed to describe interactions in solid and liquid forms of Si[18]. The potential function comprises a two-body term which describes the bond length and a three-body term which describes the bond angle, and it could be written as

$$E = \sum_i \sum_{i<j} V_2(r_{ij}) + \sum_i \sum_{i \neq j} \sum_{j<k} V_3(r_{ij}, r_{ik}, \theta_{ijk}), \quad (1)$$

where $r_{ij}$ denotes the bond length between atom *i* and atom *j*, and $\theta_{ijk}$ denotes the bond angle formed by $r_{ij}$ and $r_{ik}$, with atom *i* at the vertex. Since GULP[21] is used as the fitting tool, we further write the two terms in Eq. (1) with the form applied in GULP,

$$V_2(r_{ij}) = A_{ij}(B_{ij} r_{ij}^{-4} - 1) e^{\rho_{2ij}(r_{ij} - r_{mij})^{-1}}, \quad (2)$$

$$V_3(r_{ij}, r_{ik}, \theta_{ijk}) = K_{ijk}(\cos\theta_{ijk} - \cos\theta_{0ijk})^2 e^{\rho_{3ij}(r_{ij} - r_{mij})^{-1} + \rho_{3ik}(r_{ik} - r_{mik})^{-1}}, \quad (3)$$

where $r_{mij}$ is the cutoff distance for the interaction between atom $i$ and atom $j$, $\theta_{0ijk}$ is the equilibrium value for $\theta_{ijk}$.

If a vector composed by the candidate parameters in SW potential as $\mathbf{x} = \left[A_{ij}, B_{ij}, K_{ijk}, \rho_{2ij}, ...\right]$ is assumed, the fitting strategy in GULP is to find $\mathbf{x}_0$ that minimizes the objective function

$$F(\mathbf{x}) = \sum_{i=1}^{N_{obs}} w_i \left[f_i^{obs} - f_i^{calc}(\mathbf{x})\right]^2, \quad (4)$$

with the Broyden-Fletcher-Goldfarb-Shanno(BFGS) algorithm[22]. $f_i^{obs}$ is the $i$th observable quantity, such as lattice constant, cohesive energy, elastic constant and so on, provided by experiments or first-principles calculations. $f_i^{calc}(\mathbf{x})$ is the corresponding value calculated using the SW potential with parameter set $\mathbf{x}$. $w_i$ is the weighting factor, determined by the importance of the observable quantity $f_i^{obs}$. A lower dimension $N_{obs}$ of $\mathbf{x}$ is preferred for a simpler optimization, but lower dimension also means fewer degrees of freedom to fit the observable quantities.

In the realization on phosphorene, we first clarify the undetermined SW parameters. As is mentioned, in phosphorene there are two types of bonds, P1-P1 and P1-P2, while the P2-P2 bond is equivalent with the P1-P1 bond. Considering the two-body interaction term (see Eq. (2)), there are undetermined parameters $A_{11}, A_{12}, B_{11}, B_{12}, \rho_{211}$ and $\rho_{212}$. There are two types of bond angles, P1-P1-P1 (the first atom at the vertex) and P1-P1-P2 (equivalent with the P2-P2-P1 as marked in Figure **1**(b)). Considering the three-body interaction term (see Eq. (3)), $\theta_{0111}$ and $\theta_{0112}$ are set with the equilibrium values as 96° and 104°, respectively. $\rho_{311} = \rho_{312}$ is assumed for simplification. Thus there are three more undetermined parameters $K_{111}, K_{112}$ and $\rho_{311}$. The cutoff distances have not been treated as undetermined parameters in fitting, and we manually adjust them. $r_{m11} = r_{m12}$ is assumed according to the similar length of the two types of bond, and we finally set $r_{m11} = 2.8\text{Å}$, which is close to the average of the first and second nearest neighbor distances in phosphorene.

Next, a proper set of observable quantities is needed as the targets in fitting. What we use are lattice constants and fractional coordinates (they describe the crystal structure, which is also characterized by bond lengths and bond angles), cohesive energy ΔE and phonon frequencies. The phonon frequencies contain much information related to the thermal and mechanical properties. It is worth mentioning that the geometrical quantities should be given large weighting factors since they are generally of primary consideration. Moreover, a severe variation of lattice constants greatly varies the dimension of the first Brillouin zone (BZ), as a result, the site of a particular phonon mode in the reciprocal space shifts undesirably, making the fitting of phonon frequencies problematic. In this work, the values of observable quantities are all taken from first-principles calculations[8]. Besides the lattice constants mentioned above, the cohesive energy -3.48eV/atom, and phonon frequencies from 11 k-points (132 modes in total) in the first BZ along Y-Γ-X (Γ included, see Figure **2**) are used as observables.

The SW parameters for Si[18] are used as the initial guess. The fitting is performed for several rounds, in each following round the initial parameter set is the result of last fitting with a slight modification. Thus far, the most satisfactory parameter set discovered by us is the following: $A_{11} = 4.3807\text{eV}, A_{12} = 4.0936\text{eV}, B_{11} = 5.9563\text{Å}^4, B_{12} = 6.0042\text{Å}^4, \rho_{211} = 0.2103\text{Å}, \rho_{212} = 0.1559\text{Å}, K_{111} = 9.2660\text{eV}, K_{112} = 11.4510\text{eV}$ and $\rho_{311} = 0.4565\text{Å}$. We can also translate them into the more general form as in the original SW potential[18], and the full result is listed in Table **I** in the style of potential files in LAMMPS[23].

The lattice constants at X and Y directions of the relaxed SW phosphorene are 3.284Å and 4.590Å, very close to the first principles results 3.301Å and 4.601Å, respectively. More comparisons are listed in Table **II**. The present SW potential gives a fairly well reproduction of geometrical parameters and cohesive energy, while the acoustic velocities are generally underestimated by about 20%, which could actually be detected from the phonon dispersion comparison shown in Figure **2**. In total, the SW potential parameter set developed in the present work gives a well reproduction of the acoustic phonon branches and most of the optical phonon branches, despite of deviation due to the oversimplification in this classical potential, but which is not a serious problem here if the dominant role of acoustic phonons in thermal transport is considered. Interestingly, in the armchair direction the transverse acoustic velocity is higher than the longitudinal one, which might be related to the specially puckered structure of phosphorene.

### III. THERMAL CODUCTIVITY CALCULATION

Before applying the parameterized SW potential to calculate the thermal conductivity of phosphorene, it is necessary to test the reliability of it at nonzero temperatures, since the fitting is performed with reference at the ground state ($T = 0K$). The testing is done by thermalizing the SW phosphorene at a particular temperature and examining the structural stability. MD simulations are performed with LAMMPS[23]. We choose a phosphorene supercell of 10×10 unit cells (UCs) as the testing sample, and the crystal direction is the same as in Figure **1**(a), namely, the zigzag and armchair directions are along X and Y axes, respectively. Periodic boundary condition is applied to all directions, and thick enough vacuum layer (20Å) is adopted in the simulating box at the Z direction to avoid unwanted layer-layer interaction. The time step is set as 1.0fs in all the following MD runs.

We thermalize the sample in NPT ensemble at 1000K for $10^6$ time steps, where the strain (slight) in X and Y directions is released by varying the box size, and then run the simulation in NVE ensemble for another $10^6$ steps. Nevertheless, amorphization arises if the atomic neighbor list within the cutoff distance is frequently updated. The puckered phosphorene is flexible, especially along the armchair direction. Thus, the non-bonded atoms (beyond the first nearest neighbors) have many chances to be closer (within the potential's cutoff distance) in simulations, and amorphization arises if these atoms are treated as interaction pairs, since the SW potential tends to pull them into a wrong configuration. In fact, the similar amorphization incurred by unwanted neighbors also arises in Si described by the original SW potential, as what happened on the (001) surface of Si nanowire in MD simulations[24]. To avoid this problem, fixed atomic neighbor list is used (or only nearest neighbor interactions are considered). With this measure, the puckered structure of phosphorene is well maintained and no lattice distortion is found. It should be emphasized that the constrained neighbor list prevents the possibility in searching for other phosphorus allotropes, which should be one of the applications for MD simulation with empirical potentials. However, as a preliminary work, and our major concern is the thermal property of phosphorene, this treatment is an acceptable matter of expediency.

To characterize the structural stability, the radial distribution function g(r) is evaluated (as shown in Figure **3**(a)). The calculation of radial distribution function adopted here just considers atoms within the neighbor lists, but could effectively characterize the variation of bond lengths. The peak is well located and consistent with the equilibrium geometry of phosphorene. In addition, the time dependent kinetic energy and potential energy in the system are recorded in NVE ensemble (see Figure **3**(b)). The equipartition theorem is approximately obeyed and the total energy is conserved. Next, the thermal conductivity of phosphorene is calculated with Green-Kubo formula[25].

We follow the expression in Ref. [26],

$$\kappa_{\alpha\beta} = \frac{1}{k_B T^2 V} \int_0^{+\infty} \langle J_\alpha(t) J_\beta(0) \rangle \, dt, \qquad (5)$$

where $\kappa_{\alpha\beta}$ is the component of thermal conductivity tensor, $k_B$ is the Boltzmann constant, $T$ and $V$ are temperature and volume of the system, respectively. $J_\alpha$ is the component of heat current at $\alpha$ Cartesian direction (X, Y, and Z), and the angle brackets denote ensemble average, equivalent to time average in the MD simulations. The integral part is called heat current autocorrelation function (HCAF). The heat current is defined as

$$\mathbf{J}(t) = \frac{d}{dt}\sum_{i=1}^{N}\varepsilon_i(t)\mathbf{r}_i(t) \quad , \quad (6)$$

where $\mathbf{r}_i(t)$ is the time dependent coordinate of atom $i$ and $\varepsilon_i(t)$ is the site energy. It is calculated using LAMMPS. In its application on phosphorene, $\kappa_{XX}$ and $\kappa_{YY}$ correspond to $\kappa_{zig}$ and $\kappa_{arm}$ (subscript zig denotes zigzag and arm denotes armchair) respectively. With a phosphorene sample of $N \times N$ UCs, periodic boundary condition is applied in all directions as in the potential testing procedure. MD simulation is first run in NVT ensemble for $10^6$ steps, then switched to NVE ensemble, after another $10^6$ steps, the heat current at each time step is dumped in the next $6\times10^6$ steps. The HCAF at time $m\Delta t$ ($m = 0,1,2 \ldots$) is evaluated as

$$\langle J_\alpha(m\Delta t)J_\beta(0)\rangle = \frac{1}{N-m}\sum_{i=1}^{N-m}J_\alpha(m\Delta t + i\Delta t)J_\beta(i\Delta t) \quad , \quad (7)$$

where $\Delta t$ is the time step and $N$ is total length of HCAF. With HCAF, Eq. (5) is used to calculate the thermal conductivity, where the integration is done by the trapezoidal rule, without any fitting. One typical calculation is shown in Figure 4. Nevertheless, it is not easy to specify the converged value in Green-Kubo method, since the integrated curve is fluctuating with thermal noise. To deal with this issue, for each data point, ten independent runs are performed and averaged, each with the same upper integration limit of 200ps.

The Green-Kubo method is a formally exact approach[27] based on the fluctuation-dissipation theorem. In principle, equilibrium MD calculation has the advantage that the sample size has no limit on phonons with comparable or larger mean free paths, thus there is no finite-size effect[26]. Nevertheless, usually, there is still size effect from two competing factors[27]. As the simulation sample gets larger, low frequency phonons near Γ point, which generally make significant contributions to thermal conductivity, are excited. On the other side, the possibility of phonon-phonon scattering, which tends to decreases the thermal conductivity, is enhanced. To eliminate this size effect, the simulating sample size is gradually increased from 10×10 to 70×70 UCs (the atoms are increased from 400 to 19600) and MD simulations are performed on each. The Green-Kubo method is formally exact only in the thermodynamic limit. Thus in equilibrium molecular dynamics simulations, periodic boundary conditions are adopted with a sufficiently large supercell size. As shown in Figure 5, it is clear that while $\kappa_{arm}$ shows little size dependence, $\kappa_{zig}$ gradually increases with the increase of super cell size and saturates to a constant value after $N = 40$. Thus the saturated value is treated as the thermal conductivity of ideal 2D phosphorene with $N$ goes to infinity (thermodynamic limit). Interestingly, the in-plane thermal conductivity of phosphorene reveals strong anisotropy, the value at zigzag direction is four times larger than that at armchair direction, and this finding is consistent with other works[7-9].

The thermal conductivity of phosphorene over a wide range of temperatures is shown in Figure **6**. The thermal conductivities at both directions (zigzag and armchair) gradually decrease with the increase of temperature. They roughly obey the $T^{-1}$ law[8] but there is an obvious deviation when the temperature is high. Actually, the $T^{-1}$ law could be deduced from the BTE approach under the following condition and assumption: when the temperature is high enough and when there is only cubic anharmonicity, the specific heat is almost temperature independent and the phonon scattering rates scale linearly with temperatures[28]. Nevertheless, with a full consideration of anharmonicity in MD simulations, the deviation from this law at high temperature is not surprising. In addition to the temperature dependence, the strong anisotropy ($\kappa_{zig}/\kappa_{arm} \approx 5.0$) exists at all temperatures considered. Our calculation shows that the room temperature ($T$=300K) thermal conductivities of phosphorene at zigzag and armchair directions are $152.7 \pm 7.3$ and $33.0 \pm 2.3$ W/mK, respectively. Recently, using DFT calculation with a full solution of the BTE, Jain and McGaughey reported the strongly anisotropic in-plane thermal conductivity in monolayer phosphorene. Their values are 110 and 36 W/mK for zigzag and armchair directions, respectively [9] Thus our MD results are in considerably good agreement with the previous work. The residual differences between our MD predicted thermal conductivities and their values could be due to one or more of the following reasons. First, in principle, each phonon mode is equally excited in classical MD simulations, which is different from the quantum statistical distribution in BTE method. Second, the SW potential could not exactly reproduce the phonon dispersion predicted by first-principles calculations, which directly influences the phonon group velocities and the phonon-phonon scattering rates. Third, in BTE calculations, the translational invariance and truncation at third-order force constants also significantly affect the calculated value of thermal conductivity [8, 9]. On the whole, the MD simulations provide fairly good agreement with first-principles based BTE results, and the most important characteristic in phosphorene, the strong anisotropic thermal conductivity, is well revealed by MD calculations with the parameterized empirical potential.

## IV. PHONON INFORMATION

We have calculated the thermal conductivity of phosphorene. Nevertheless, the Green-Kubo method does not provide any mode-wise information, such as phonon group velocities, lifetimes and mean free paths (MFPs), which are commonly of interest. In this section, we use lattice dynamics calculations and MD simulations to evaluate these quantities in phosphorene, which are actually based on the relaxation time approximation (RTA). These quantities could explain the strong anisotropy of thermal conductivity in phosphorene.

With the RTA, the thermal conductivity may be calculated as

$$\kappa_{\alpha\alpha} = \sum_{\mathbf{k},\nu} c\binom{\mathbf{k}}{\nu} v_\alpha^2 \binom{\mathbf{k}}{\nu} \tau\binom{\mathbf{k}}{\nu} , \qquad (8)$$

where $c\binom{\mathbf{k}}{\nu}$ is the specific heat contributed by a phonon mode with wave vector **k** at dispersion branch $\nu$, equal to $k_B/V$ in the classical limit in MD simulations, $v_\alpha$ is the component of group velocity at $\alpha$ Cartesian direction, and $\tau$ is phonon lifetime (or relaxation time). For phosphorene, only those modes at Γ-X and Γ-Y (see Figure **2**) in the first BZ are considered, since they are the most representative ones for zigzag and armchair directions, respectively. The dimension of sample used for phonon lifetime analysis is 40×40, thus, the allowed wave vectors are $(m\pi/20a, 0,0)$ and $(0, n\pi/20b, 0)$, where $m$ and $n$ are integers from 1 to 20 (half first BZ), $a$ and $b$ are lattice constants at zigzag and armchair directions, respectively. The phonon group velocities are calculated with a backward difference technique on the phonon dispersion calculated with GULP. The phonon lifetime is evaluated with frequency-domain normal mode decomposition[29] as following.

We project the atomic velocities into reciprocal space as

$$\dot{q}\binom{\mathbf{k}}{\nu}; t = \sum_{j,l} \sqrt{\frac{m_j}{N}} e^{-i\mathbf{k}\cdot\mathbf{r_o}(lj)} \boldsymbol{\varepsilon}_j^*\binom{\mathbf{k}}{\nu} \cdot \dot{\mathbf{u}}\binom{l}{j}; t , \qquad (9)$$

where $\dot{q}$ is normal mode velocity, $\mathbf{r_o}(lj)$ and $\dot{\mathbf{u}}\binom{l}{j}; t$ are the equilibrium position and time dependent velocity of the $j$th atom in the $l$th UC, $\boldsymbol{\varepsilon}_j^*$ is the complex conjugate of the eigenvector associated with the $j$th atom at a given mode, $m_j$ is atomic mass and $N$ is the total number of UCs. Then the power spectrum of $\dot{q}\binom{\mathbf{k}}{\nu}; t$ is evaluated with Fourier transformation,

$$\Phi\binom{\mathbf{k}}{\nu}, f = \int_{-\infty}^{+\infty} \langle \dot{q}^*\binom{\mathbf{k}}{\nu}; 0 \dot{q}\binom{\mathbf{k}}{\nu}; t \rangle e^{i2\pi f t} dt . \qquad (10)$$

A Lorentz fitting of $\Phi\left(\begin{smallmatrix}\mathbf{k}\\\nu\end{smallmatrix}, f\right)$ is performed in the neighborhood of the harmonic eigen frequency $f_0\left(\begin{smallmatrix}\mathbf{k}\\\nu\end{smallmatrix}\right)$ as

$$\Phi\left(\begin{smallmatrix}\mathbf{k}\\\nu\end{smallmatrix}, f\right) = \frac{C\left(\begin{smallmatrix}\mathbf{k}\\\nu\end{smallmatrix}\right)}{4\pi^2\left(f - f_a\left(\begin{smallmatrix}\mathbf{k}\\\nu\end{smallmatrix}\right)\right)^2 + \Gamma^2\left(\begin{smallmatrix}\mathbf{k}\\\nu\end{smallmatrix}\right)}, \qquad (11)$$

where $C$, $f_a$ and $\Gamma$ are unknown parameters. $f_a$ is the anharmonic eigen frequency at non-zero temperature, usually with a small shift from $f_0$. The phonon lifetime $\tau\left(\begin{smallmatrix}\mathbf{k}\\\nu\end{smallmatrix}\right)$ is equal to $1/2\Gamma\left(\begin{smallmatrix}\mathbf{k}\\\nu\end{smallmatrix}\right)$, inverse to the width of the Lorentz peak. The atomic velocities are recorded every 20 time steps (0.02ps) in the NVE ensemble at 300K, and six independent runs are performed to minimize the uncertainty. One typical calculation is illustrated in Figure **7**.

The frequency dependent phonon group velocities and lifetimes are shown in Figure **8**. Generally, the phonon group velocities at zigzag direction ($\Gamma$-X) are higher than those of phonons at armchair ($\Gamma$-Y) direction with similar frequencies, this is especially obvious for acoustic phonons, which make dominant contributions to the thermal transport as already proved[7-9]. In contrast, phonon lifetimes at the two directions are very close to each other except that the lifetimes at zigzag direction are a little longer than those in the armchair direction within 2~5THz and the optical phonon part. Therefore, according to Eq. (8), we speculate that the remarkably directional dependence of thermal conductivity is mainly due to the anisotropy of phonon group velocities. A rough estimation with the acoustic velocity ratio (about two) and Eq. (8) reveals a thermal conductivity ratio of four at the two directions, which is close to the ratio of five predicted by Green-Kubo method.

We could evaluate the mode dependent phonon MFP as $\left|\mathbf{v}\left(\begin{smallmatrix}\mathbf{k}\\\nu\end{smallmatrix}\right)\right|\tau\left(\begin{smallmatrix}\mathbf{k}\\\nu\end{smallmatrix}\right)$. Whereas, an effective MFP is usually of more interest since it is useful when designing the functional

nanostructures. For instance, it could be used to judge at which length scale the wave effect of phonons is important, so that the phononic crystal[30] with a proper period length efficiently works. The effective MFP is defined as [31]:

$$\lambda = \frac{\kappa}{\sigma} \qquad (12)$$

where κ is the intrinsic thermal conductivity of material at diffusive transport region, $\sigma$ is the thermal conductance at the ballistic transport limit, and $\lambda$ is the so called effective MFP. Using first-principles calculations and NEGF method, thermal conductance and MFP of MoS$_2$ sheet [31] have been studied, and this formula has been used to evaluate the value of thermal conductivity. To evaluate $\lambda$ of phosphorene, we make use of the intrinsic thermal conductivity κ from Green-Kubo method in this work and the ballistic thermal conductance $\sigma$ in our previous work[7]. At 300K, $\kappa_{zig} = 152.7$ W/mK, $\kappa_{arm} = 33.0$ W/m, $\sigma_{zig} = 1.08 \times 10^9$ W/m$^2$K and $\sigma_{arm} = 0.76 \times 10^9$ W/m$^2$K. Thus, the effective MFPs are about 141.4 and 43.4nm at zigzag and armchair directions, respectively. The MFPs of phosphorene are one order of magnitude lower than that of graphene[32], while one order higher than that of MoS$_2$ [33].

## V. DSICUSSION AND SUMMARY

Although satisfying descriptions of the crystal structure, cohesive energy, phonon dispersion and thermal conductivity of monolayer black phosphorene are reproduced, it must be mentioned that this is a preliminary work on phosphorene from such an approach, despite of some deficiencies. The most significant limitation is related to the reliability of the set of potential parameters fitted in the present work. First, this interatomic potential is specified for black phosphorene. The enumeration of phosphorous atoms, bond lengths and bond angles (see Table I) may limit its application to other phosphorus allotropes such as the blue phosphorus [34]. A more widely transferable empirical interatomic potential might call for a more complex form, such as the Tersoff [35], Brenner [36] and EDIP [37] types. Besides, a long range force between non-bonded atoms might also need to be considered. Furthermore, the fixed atomic neighbor list adopted in our calculation influences other possible applications, such as modeling melting, deformation and fracture. The amorphization phenomenon predicted by this potential when all neighbor pairs are treated shows that the underlying description may be incomplete. All of these show that further investigations are deserved with this potential as a starting point.

In summary, we have parameterized a set of Stilling-Weber empirical potential parameters for monolayer black phosphorene, and evaluated the thermal conductivity by using equilibrium molecular dynamics simulations. Unlike the isotropic in-plane thermal conductivity in graphene and $MoS_2$, thermal transport in monolayer black phosphorene exists a strongly directional dependence, with the values of room temperature thermal conductivity along zigzag and armchair directions are 152.7 and 33.0 W/mK, respectively. The MD predicted values are close with results from first-principles and BTE calculations.

# References

[1] H. Liu, A. T. Neal, and P. D. Ye, ACS nano **6**, 8563 (2012).
[2] L. Li, Y. Yu, G. J. Ye, Q. Ge, X. Ou, H. Wu, D. Feng, X. H. Chen, and Y. Zhang, Nat Nano **9**, 372 (2014).
[3] H. Liu, A. T. Neal, Z. Zhu, Z. Luo, X. Xu, D. Tománek, and P. D. Ye, ACS Nano **8**, 4033 (2014).
[4] N. Yang, X. Xu, G. Zhang, and B. Li, AIP Advances **2**, 041410 (2012).
[5] R. Fei, A. Faghaninia, R. Soklaski, J.-A. Yan, C. Lo, and L. Yang, Nano Letters **14**, 6393 (2014).
[6] J. Zhang, H. J. Liu, L. Cheng, J. Wei, J. H. Liang, D. D. Fan, J. Shi, X. F. Tang, and Q. J. Zhang, Sci. Rep. **4** (2014).
[7] Z.-Y. Ong, Y. Cai, G. Zhang, and Y.-W. Zhang, The Journal of Physical Chemistry C **118**, 25272 (2014).
[8] L. Zhu, G. Zhang, and B. Li, Physical Review B **90**, 214302 (2014).
[9] A. Jain and A. J. H. McGaughey, Sci. Rep. **5**, 8501 (2015).
[10] W. Xu, G. Zhang, and B. Li, Journal of Applied Physics **116**, 134303 (2014).
[11] R. Rurali, X. Cartoixà, and L. Colombo, Physical Review B **90**, 041408 (2014).
[12] N.-N. Li, Z.-D. Sha, Q.-X. Pei, and Y.-W. Zhang, The Journal of Physical Chemistry C **118**, 13769 (2014).
[13] Z. D. Sha, S. S. Quek, Q. X. Pei, Z. S. Liu, T. J. Wang, V. B. Shenoy, and Y. W. Zhang, Sci. Rep. **4**, 5991 (2014).
[14] G. Zhang, *Nanoscale Energy Transport and Harvesting: A Computational Study* (Pan Stanford, 2014).
[15] B. Qiu and X. Ruan, Physical Review B **80**, 165203 (2009).
[16] C. Sevik, A. Kinaci, J. B. Haskins, and T. Çağın, Physical Review B **84**, 085409 (2011).
[17] X. Zhang, H. Xie, M. Hu, H. Bao, S. Yue, G. Qin, and G. Su, Physical Review B **89**, 054310 (2014).
[18] F. H. Stillinger and T. A. Weber, Physical Review B **31**, 5262 (1985).
[19] X. W. Zhou, D. K. Ward, J. E. Martin, F. B. van Swol, J. L. Cruz-Campa, and D. Zubia, Physical Review B **88**, 085309 (2013).
[20] J. Chen, G. Zhang, and B. Li, The Journal of Chemical Physics **135**, 104508 (2011).
[21] J. D. Gale and A. L. Rohl, Molecular Simulation **29**, 291 (2003).
[22] D. F. Shanno, Mathematics of computation **24**, 647 (1970).
[23] S. Plimpton, Journal of computational physics **117**, 1 (1995).
[24] S. G. Volz and G. Chen, Applied Physics Letters **75**, 2056 (1999).
[25] R. Kubo, Journal of the Physical Society of Japan **12**, 570 (1957).
[26] P. K. Schelling, S. R. Phillpot, and P. Keblinski, Physical Review B **65**, 144306 (2002).
[27] A. J. Ladd, B. Moran, and W. G. Hoover, Physical Review B **34**, 5058 (1986).
[28] S. Bickham and J. Feldman, Physical Review B **57**, 12234 (1998).
[29] A. J. H. McGaughey and J. M. Larkin, Annual Review of Heat Transfer **17**, 49 (2014).
[30] L. Yang, N. Yang, and B. Li, Nano letters **14**, 1734 (2014).
[31] Y. Cai, J. Lan, G. Zhang, and Y.-W. Zhang, Physical Review B **89**, 035438 (2014).
[32] S. Ghosh, W. Bao, D. L. Nika, S. Subrina, E. P. Pokatilov, C. N. Lau, and A. A. Balandin, Nat Mater **9**, 555 (2010).
[33] X. Liu, G. Zhang, Q.-X. Pei, and Y.-W. Zhang, Applied Physics Letters **103**, 133113 (2013).
[34] Z. Zhu and D. Tománek, Phys. Rev. Lett. **112**, 176802 (2014).
[35] J. Tersoff, Physical Review B **39**, 5566 (1989).
[36] D. W. Brenner, Physical Review B **42**, 9458 (1990).
[37] M. Z. Bazant and E. Kaxiras, Physical Review Letters **77**, 4370 (1996).

Table I. SW potential parameters for phosphorene.

|  | $\varepsilon$ (eV) | $\sigma$ (Å) | $a$ | $\lambda$ | $\gamma$ | $\cos\theta_0$ | $A$ | $B$ | $p$ | $q$ | tol |
|---|---|---|---|---|---|---|---|---|---|---|---|
| **P1P1P1** | 1.0 | 0.2103 | 13.3143 | 9.2660 | 2.1707 | -0.1045 | 4.3807 | 3045.2 | 4.0 | 0 | 0 |
| **P2P2P2** | 1.0 | 0.2103 | 13.3143 | 9.2660 | 2.1707 | -0.1045 | 4.3807 | 3045.2 | 4.0 | 0 | 0 |
| **P1P2P2** | 1.0 | 0.1559 | 17.9602 | 0 | 2.9282 | 0 | 4.0936 | 10164.1 | 4.0 | 0 | 0 |
| **P2P1P1** | 1.0 | 0.1559 | 17.9602 | 0 | 2.9282 | 0 | 4.0936 | 10164.1 | 4.0 | 0 | 0 |
| **P1P1P2** | 1.0 | 0 | 0 | 11.4510 | 0 | -0.2419 | 0 | 0 | 0 | 0 | 0 |
| **P1P2P1** | 1.0 | 0 | 0 | 11.4510 | 0 | -0.2419 | 0 | 0 | 0 | 0 | 0 |
| **P2P1P2** | 1.0 | 0 | 0 | 11.4510 | 0 | -0.2419 | 0 | 0 | 0 | 0 | 0 |
| **P2P2P2** | 1.0 | 0 | 0 | 11.4510 | 0 | -0.2419 | 0 | 0 | 0 | 0 | 0 |

Table II. Geometrical parameters, cohesive energy, and acoustic velocities predicted by first-principles calculations as compared to those predicted by SW potential. Values with * are cited from Ref. [34].

|  | $l_{in}$ (Å) | $l_{out}$ (Å) | $\theta_{in}$ (°) | $\theta_{out}$ (°) | $\Delta E$ (eV/atom) | $v_{LA}^{zig}$ (Km/s) | $v_{TA}^{zig}$ (Km/s) | $v_{LA}^{arm}$ (Km/s) | $v_{TA}^{arm}$ (Km/s) |
|---|---|---|---|---|---|---|---|---|---|
| **1st principles** | 2.221 | 2.259 | 96.001 | 103.961 | -3.48 | 8.57 (7.8*) | 4.51 | 4.30 | 4.45 (3.8*) |
| **SW potential** | 2.210 | 2.258 | 95.999 | 104.000 | -3.48 | 6.43 | 3.50 | 3.46 | 3.50 |

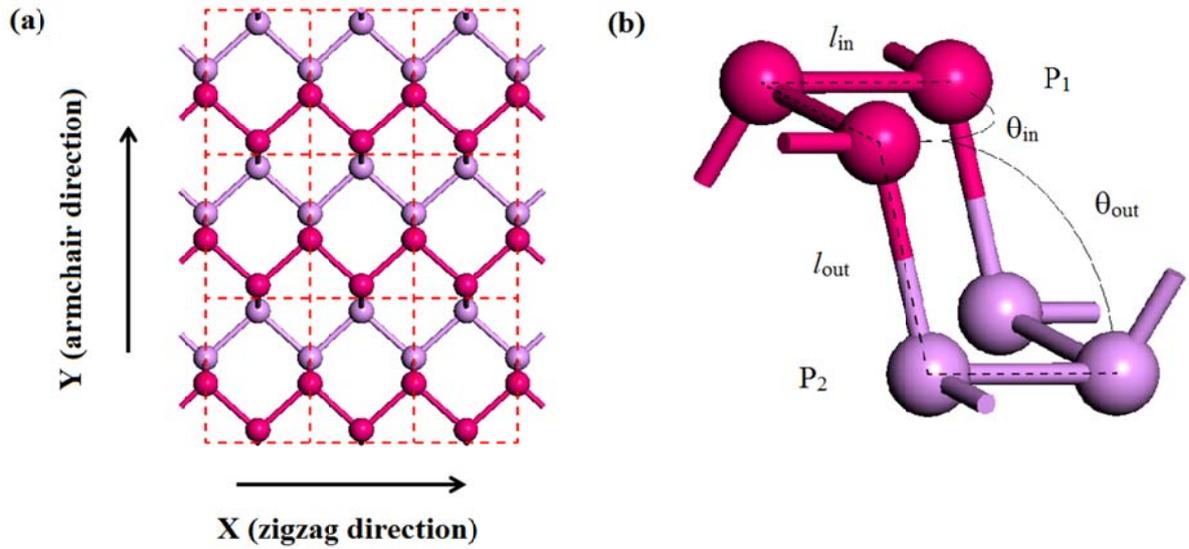

Figure 1. Schematic of phosphorene. (a) Top view of the 3×3 supercell, the red dashed lines indicate the unit cells, the zigzag and armchair directions are arranged along X and Y axes respectively. (b) Enlarged drawing of one unit cell (including full boundary atoms), the top three (dark-coloured) and bottom three (light-coloured) phosphorus atoms are enumerated as P1 and P2 respectively.

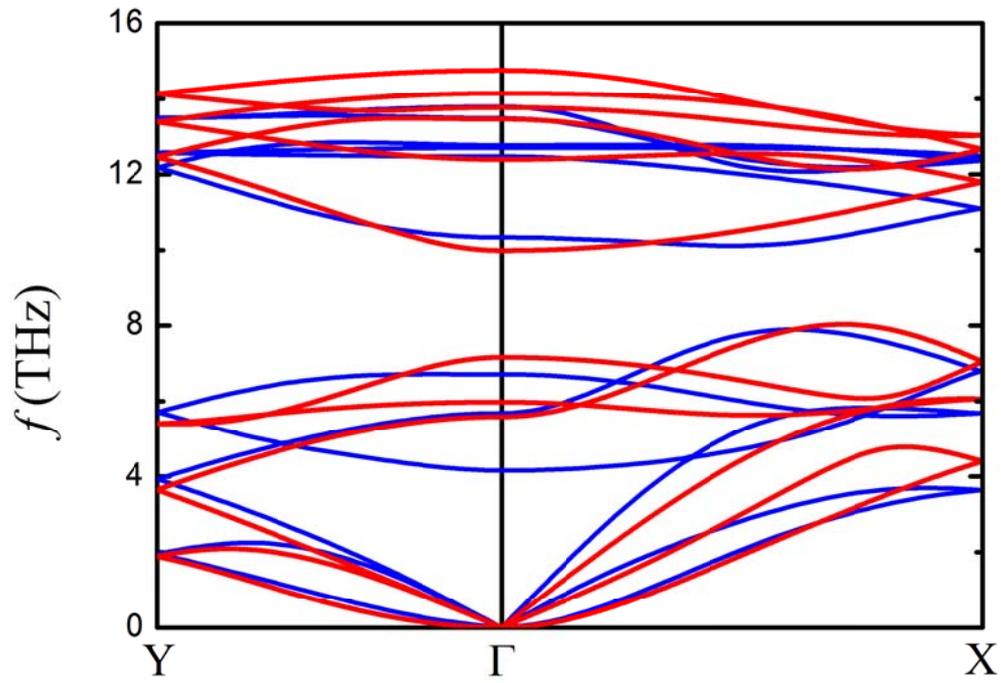

Figure 2 Phonon dispersions of phosphorene along Y-Γ-X in the first BZ. The blue and red curves are obtained by first-principles calculations and the parameterized SW potential respectively.

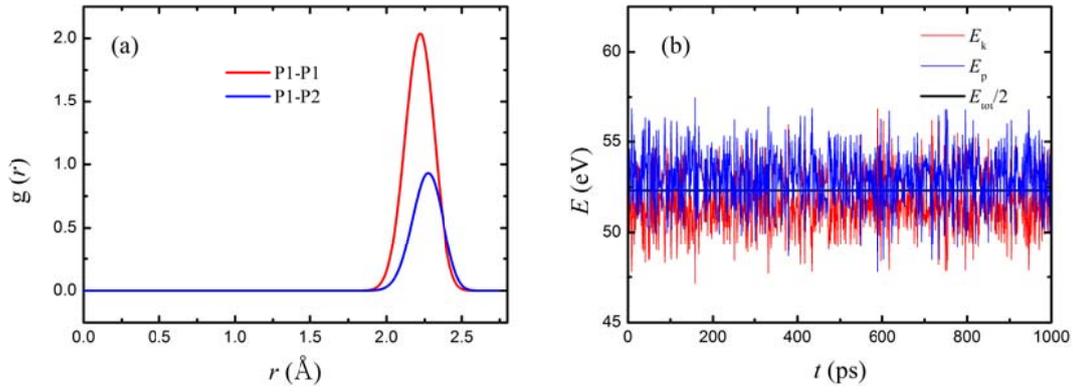

Figure 3. SW phosphorene at 1000K. (a) Radial distribution function, P1-P1 means the statistics of P1 atoms around P1 atoms, and P1-P2 means statistics of P2 atoms around P1 atoms. (b) The time dependent kinetic energy $E_k$, potential energy $E_p$, and half of total energy $E_{tot}$ of the sample, $E_p$ and $E_{tot}$ have been subtracted with their ground state values (when $T$=0K).

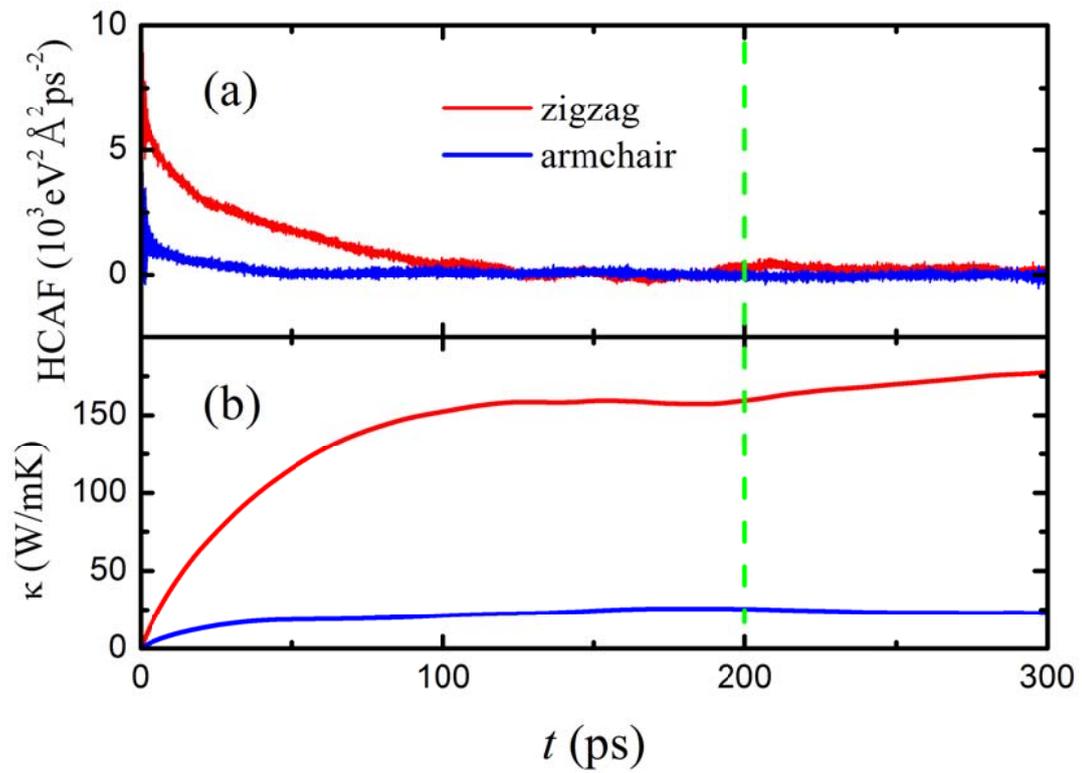

Figure 4. Evaluating thermal conductivity of phosphorene at 300K with Green-Kubo method, the simulating sample is of 60×60 UCs. The green dashed line indicates the integration time limit. (a) The HCAF. (b) The dependence of integration result on the upper limit.

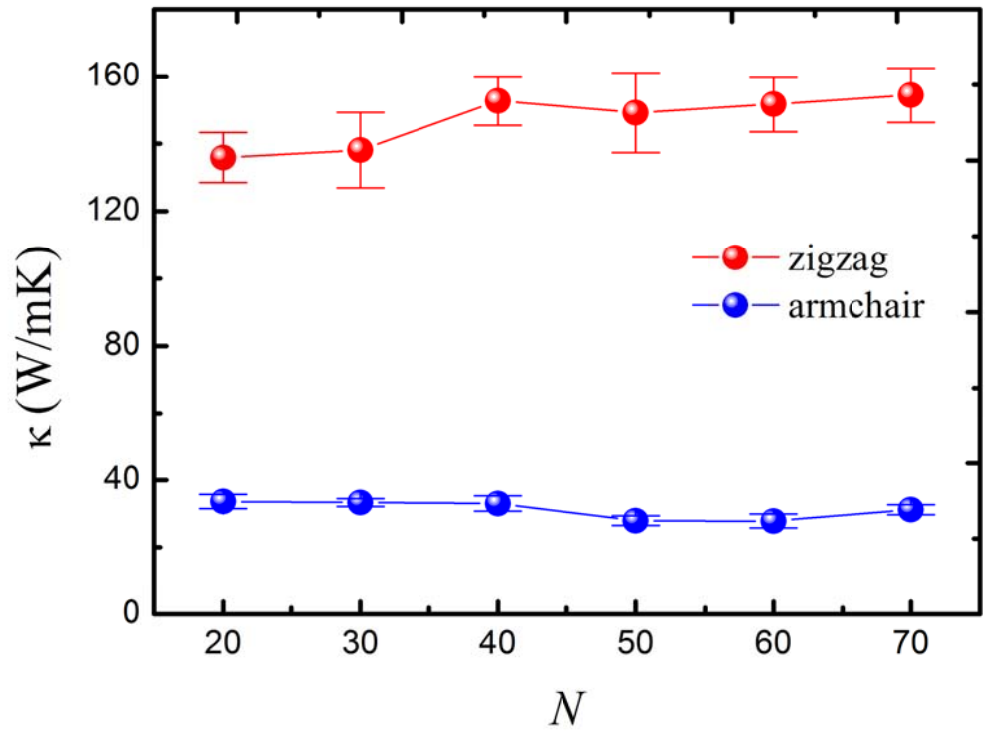

Figure 5. Thermal conductivity of phosphorene at 300K calculated with samples of different sizes (**N** × **N** UCs).

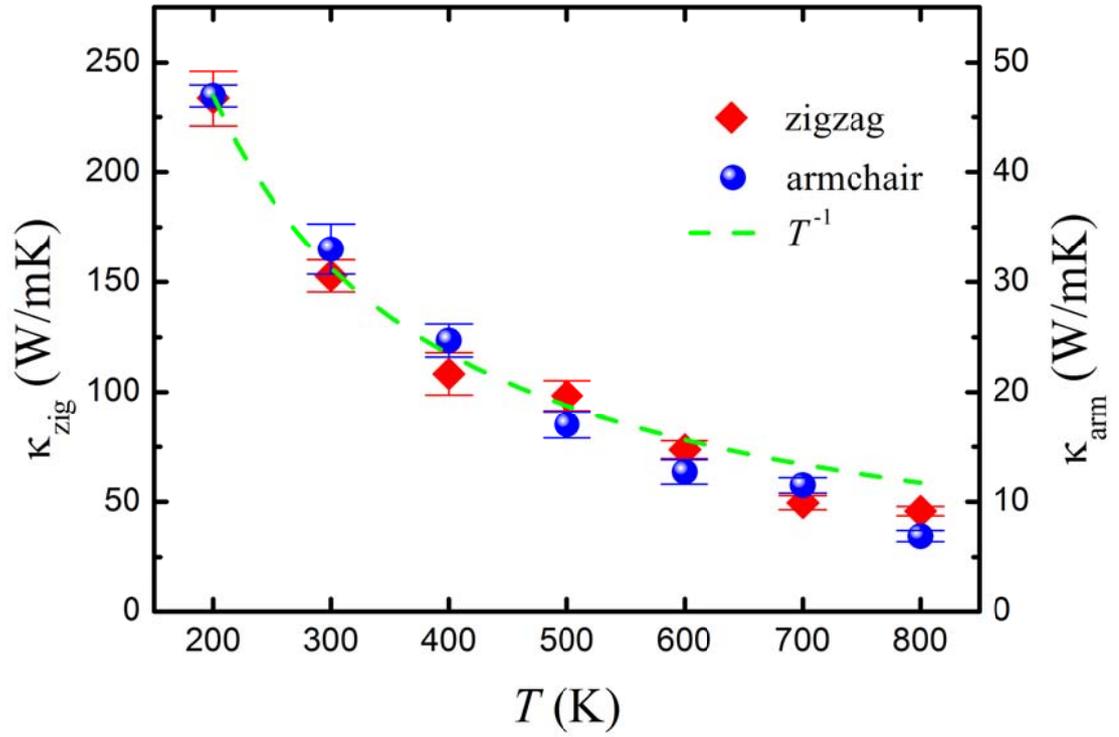

Figure 6. Temperature dependent thermal conductivity of phosphorene. Values of $\kappa_{zig}$ and $\kappa_{arm}$ are scaled on the left and right vertical axes, respectively, and they have a ratio of five. The green dashed line denotes an extrapolation according to the $T^{-1}$ law.

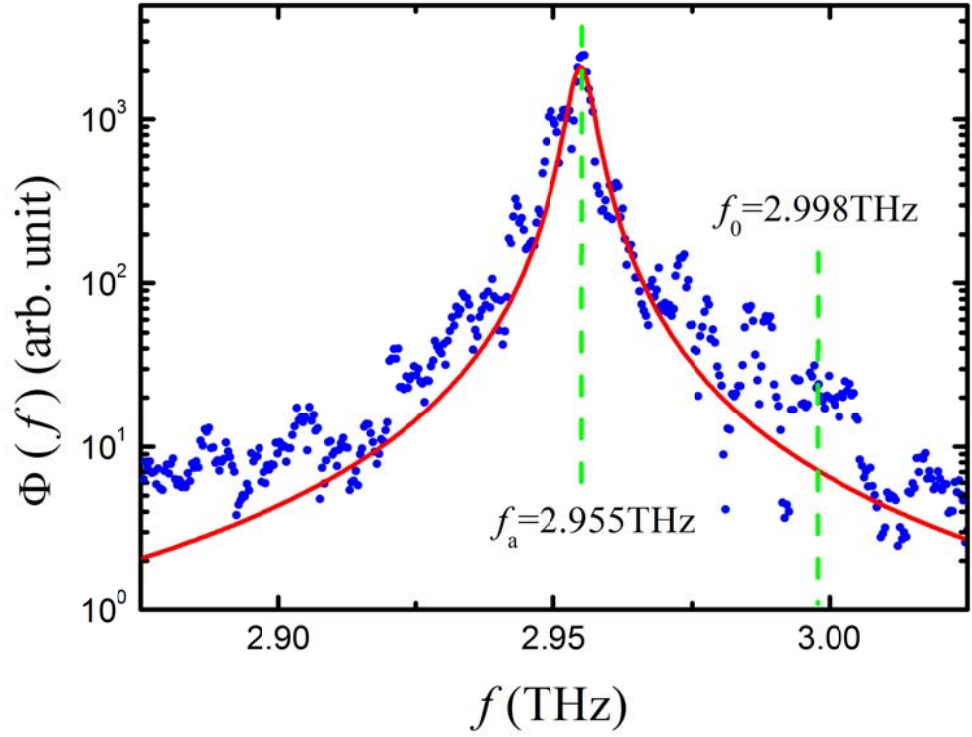

Figure 7. Power spectrum (blue dots) of a transverse acoustic phonon mode, whose wave vector and harmonic eigen frequency are ($\pi/2a, 0, 0$) and 2.998THz, respectively, and the Lorentz fitting (red curves), which shows that the eigen frequency is shifted to 2.955THz at 300K. The phonon lifetime is evaluated to be 31.7ps.

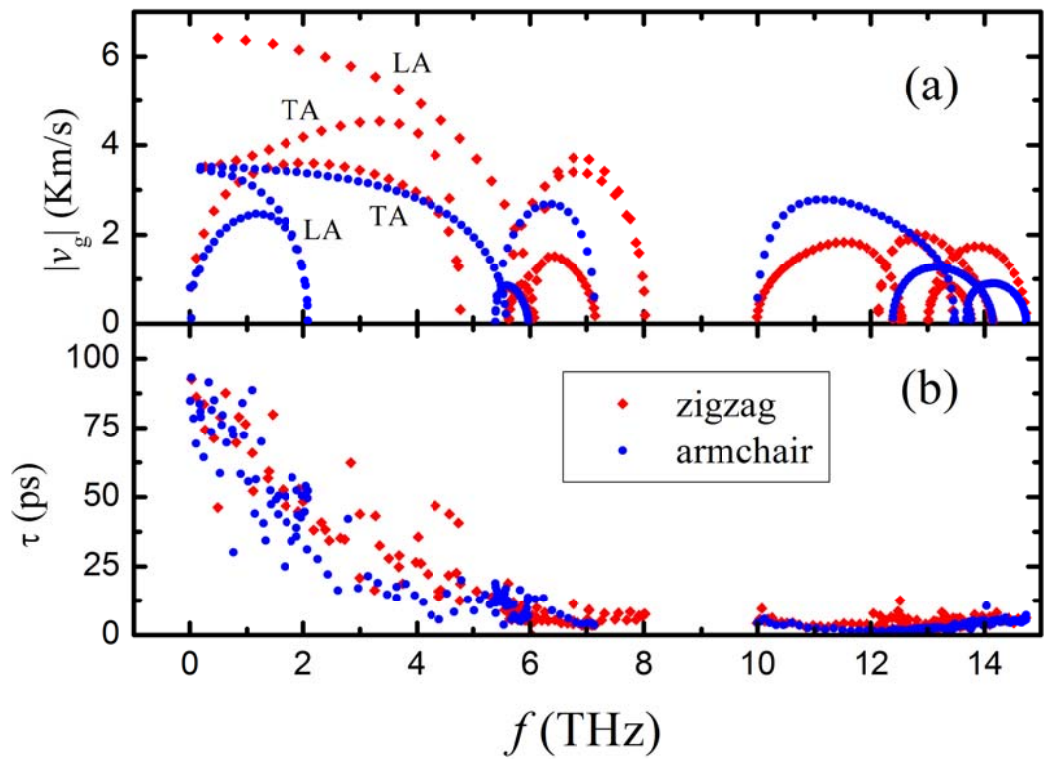

Figure 8. Phonon group velocities and lifetimes at zigzag and armchair directions. (a) Phonon group velocities. (b) Phonon lifetimes.